\documentclass[conference]{IEEEtran}
\IEEEoverridecommandlockouts

\usepackage[bookmarks=false, draft]{hyperref}

 \usepackage{cite}

\usepackage{amsmath,amssymb,amsfonts}
\usepackage{algpseudocode} 
\usepackage{algorithm}     
\usepackage{graphicx}
\usepackage{textcomp}
\usepackage{color}
\usepackage{lipsum} 
\usepackage{xurl}

\usepackage{booktabs} 
\usepackage{array} 
\usepackage{makecell} 
\usepackage{url}
 \usepackage{booktabs}
 \usepackage{amssymb}
\usepackage{tabularx}
\usepackage{optidef}
 \usepackage{graphicx}
\usepackage{wrapfig}
\usepackage{afterpage}
\usepackage{float}
\usepackage{multirow}
 \usepackage{array}
\usepackage{hhline}
 \usepackage{tabularx}
 \usepackage{makecell}
\usepackage{indentfirst}
\usepackage{adjustbox}
 \usepackage{amsmath}
 \usepackage{caption}
\usepackage{soul}
\usepackage{soulpos}

\usepackage{bm} 
\usepackage{mathptmx} 

\usepackage{ifthen} 
\usepackage{xcolor} 

\newcommand{\showchanges}{0}

\newcommand{\added}[1]{\ifthenelse{\equal{\showchanges}{1}}{\textcolor{blue}{#1}}{#1}}
\newcommand{\deleted}[1]{\ifthenelse{\equal{\showchanges}{1}}{\textcolor{red}{\sout{#1}}}{}}

\begin{document}


\title{Intrusion Detection on Resource-Constrained IoT Devices with Hardware-Aware ML and DL}

\author{
\IEEEauthorblockN{ 
A. Diab$^{1}$,
A. Chehade$^{2}$,
E. Ragusa$^{2}$,
P. Gastaldo$^{2}$,
R. Zunino$^{2}$,
A. Baghdadi$^{3}$,
M. Rizk$^{1,4}$
}
\IEEEauthorblockA{
\textit{$^{\text{1}}$ Faculty of Sciences, Lebanese University, Hadath, Lebanon}\\
\textit{$^{\text{2}}$ DITEN, University of Genova, Genova, Italy}\\
\textit{$^{3}$IMT Atlantique, Lab-STICC UMR CNRS 6285, Brest, France}\\
\textit{$^{\text{4}}$ School of Arts and Sciences, Lebanese American University, Beirut, Lebanon}
}
}

\vspace{-5cm}
\maketitle



\begin{abstract}
This paper proposes a hardware-aware intrusion detection system (IDS) for Internet of Things (IoT) and Industrial IoT (IIoT) networks; it targets scenarios where classification is essential for fast, privacy-preserving, and resource-efficient threat detection. The goal is to optimize both tree-based machine learning (ML) models and compact deep neural networks (DNNs) within strict edge-device constraints. This allows for a fair comparison and reveals trade-offs between model families. We apply constrained grid search for tree-based classifiers and hardware-aware neural architecture search (HW-NAS) for 1D convolutional neural networks (1D-CNNs). Evaluation on the Edge-IIoTset benchmark shows that selected models meet tight flash, RAM, and compute limits: LightGBM achieves 95.3\% accuracy using 75\,KB flash and 1.2\,K operations, while the HW-NAS–optimized CNN reaches 97.2\% with 190\,KB flash and 840\,K floating-point operations (FLOPs). We deploy the full pipeline on a Raspberry~Pi~3~B+, confirming that tree-based models operate within 30\,ms and that CNNs remain suitable when accuracy outweighs latency. These results highlight the practicality of hardware-constrained model design for real-time IDS at the edge.
\end{abstract}

\IEEEoverridecommandlockouts
\begin{IEEEkeywords}
Intrusion detection, Edge AI, Hardware-aware NAS, Traffic classification, Lightweight deep learning, TinyML.
\end{IEEEkeywords}

\vspace{-0.2cm}
\IEEEpeerreviewmaketitle

\section{Introduction}

The widespread deployment of Internet of Things (IoT) systems has expanded the attack surface of modern networks, which now include critical infrastructure and operational environments vulnerable to advanced cyber threats \cite{paper2,paper7}. In such contexts, intrusion detection systems (IDS) must deliver high detection performance while operating within tight hardware constraints. Cloud-based IDS often add latency, rely on constant connectivity, and raise privacy issues due to traffic offloading to remote servers \cite{paper25}. These limitations have motivated a shift toward local inference, where detection is performed directly at the edge. This work addresses the design of IDS models that operate under constrained memory, compute, and energy budgets, while ensuring timely and accurate predictions in resource-limited environments.

Several recent studies have applied machine learning (ML) and deep learning (DL) to traffic-based intrusion detection, reporting satisfactory accuracy on public benchmarks \cite{paper4,paper6,ullah2024attention,hamza2023malware}; however, most ignore edge constraints such as memory, computation, and latency, causing many models that perform well offline to exceed hardware budgets when deployed.



This work focuses on tabular traffic data, where each flow is represented by handcrafted features from packet headers and metadata, capturing protocol and transport-level behavior. These features must be extractable in real time without delay or heavy overhead. Tree-based ML models, such as decision tree ensembles, are generally efficient but still require tuning to meet memory and latency constraints. Their performance can vary with hyperparameter configurations \cite{paper26}. Deep models, such as convolutional and recurrent neural networks, often demand structural adaptation to operate within the compute and memory limits of embedded platforms \cite{paper25}. Supporting both families calls for model design strategies that balance detection performance with hardware feasibility.

Hardware-aware neural architecture search (HW-NAS) has recently been employed across domains to design efficient DL models \cite{paper20,paper27}. 
In this work, we frame model selection as a constrained optimization problem, aiming to maximize accuracy under strict memory and compute limits. For DL, we adopt HW-NAS to generate compact 1D convolutional neural networks (1D-CNNs) tailored to tabular traffic data, with memory and computation constraints embedded into the search. For tree-based ML models, we apply a constrained grid search under the same hardware considerations. We characterize these constraints in terms of available flash, RAM, and compute load; the latter is measured using operation counts (Ops) for ML models, which rely on integer comparisons and logical branching, and floating-point operations (FLOPs) for CNNs, which perform dense arithmetic. This dual-track strategy enables a fair, hardware-constrained comparison between ML and DL approaches for edge-based IDS.


The main contributions of this study are as follows:
\begin{itemize}


\item We formulate IDS for IoT as a constrained optimization problem and address it using two strategies: grid search for tree-based ML and HW-NAS for CNNs.


\item We present a systematic comparison of the models derived from both strategies, all designed under shared hardware constraints, and tailored for deployment on IoT gateway platforms.

\item Our hardware-aware search produces highly efficient yet accurate models: LightGBM achieves 95.3\% accuracy using just 75\,KB of flash and 1.2\,K operations, while the optimized CNN reaches 97.2\% accuracy within 190\,KB and 840\,K FLOPs. The latter is up to 10$\times$ more efficient than recent DL baselines while preserving accuracy.

\item We demonstrate full deployment on a Raspberry~Pi~3~B+, where feature extraction completes in under 1\,ms, tree-based models deliver sub-30\,ms inference, and CNNs remain viable when accuracy is prioritized over latency.
\end{itemize}

The rest of the paper is organized as follows. 
Section~\ref{RelatedWork} reviews related work.
Section~\ref{Method} describes the adopted methodology.
Section~\ref{Results} presents the results and discussions.
Section~\ref{Deployment} 
presents the hardware implementation and deployment results.
Finally, 
Section~\ref{Conclusion} concludes the paper and outlines future work.

\vspace{-0.1cm}
\section{Related Work}
\label{RelatedWork}


\subsection{ML and DL Approaches}

Extensive research has explored intrusion detection using traditional ML methods such as decision trees (DT) and random forests (RF), as well as DL architectures like convolutional neural networks (CNNs) and recurrent neural networks (RNNs). These approaches have demonstrated strong performance across various datasets.


The study in \cite{paper2} reported that RF achieved the highest performance with an accuracy of 80.83\%, while \cite{hamza2023malware} evaluated several ML models for malware detection and found RF to be the top performer with 94.1\% on the Edge-IIoTset dataset~\cite{paper2}.



Several DL-based IDS models have also been proposed. The work in~\cite{paper6} has developed a Long Short-Term Memory (LSTM) network with three hidden layers, achieving 86.6\% accuracy. In \cite{paper28}, the authors have evaluated several tree-based models on a real-world traffic dataset, where XGBoost achieved the best performance with 89.09\% accuracy using feature selection. While DL models show strong accuracy, many require optimization to meet the memory and compute constraints of edge deployment.


Other works have explored hybrid and ensemble approaches. 
The study in \cite{de2022hybrid} has proposed a hybrid CNN-LSTM model for privacy-aware intrusion detection, achieving 97.14\% accuracy. In \cite{ullah2024attention}, the authors have proposed an attention-based hybrid model (AB-HDL) combining CNN and autoencoder with an attention mechanism, achieving 99.95\% accuracy.



These approaches improve detection accuracy but often overlook hardware constraints such as memory, latency, and energy, which are critical for practical deployment on resource-constrained devices.

\subsection{Hardware-Aware Optimization for ML and DNN}

Model optimization is essential for meeting performance and deployment goals, especially on resource-constrained devices. A popular approach for optimizing deep models is neural architecture search (NAS)~\cite{paper19,paper20}, which automates the design of deep neural networks (DNNs) to maximize accuracy. To support real-world deployment, HW-NAS  extends this idea by including constraints on memory, compute, and energy in the search process.


Despite this progress, many NAS and optimization approaches still focus mainly on accuracy~\cite{paper20,paper26}, overlooking practical deployment constraints. Traditional hyperparameter tuning and heuristics also rarely account for strict resource budgets, which limits their use on embedded platforms~\cite{paper26}.
Recent HW-NAS methods address this by generating models for limited devices~\cite{paper24,paper25,paper22,paper23}, and have shown success in domains where performance and efficiency must be balanced.




This work explores hardware-aware optimization for IDS, as prior efforts often focus only on accuracy. We aim to find models that stay accurate while fitting the resource limits of edge devices. To achieve this, we apply optimization to both DNN and traditional ML models to design efficient solutions and expose their tradeoffs under constrained settings.

\section{Methodology}
\label{Method}
\vspace{-0.1cm}


Deploying ML and DNN models on constrained platforms requires optimizing predictive performance under strict hardware budgets. Our proposal addresses this challenge using two strategies: tree-based ML models selected via grid search, and compact DNNs designed through HW-NAS. We formalize model selection as an optimization problem, where $m$ denotes a candidate model and $\mathcal{M}$ the set of all candidate models under consideration:

\begin{maxi}|s|
    {m \in \mathcal{M}} {\mathrm{Accuracy}_{\mathrm{val}}(m)}
    {}{}
    \addConstraint{\mathrm{Flash}(m) \le F_{\max}}
    \addConstraint{\mathrm{RAM}(m) \le T_{\max}}
    \addConstraint{\mathrm{Compute}(m) \le C_{\max}}
\end{maxi}

Here, $\mathrm{Flash}(m)$ denotes the non-volatile memory to store the model (e.g., all decision nodes for tree ensembles or weights for DNNs), $\mathrm{RAM}(m)$ is the peak intermediate memory usage during inference, and $\mathrm{Compute}(m)$ quantifies the per-inference computational cost. For DNNs, $\mathrm{Compute}(m)$ is expressed in floating-point operations (FLOPs), while for tree-based ML models it is expressed in decision operations (Ops). The thresholds $(F_{\max}, T_{\max}, C_{\max})$ follow target device specifications and, as in prior work~\cite{paper22,paper23,paper24,paper25}, ensure the resulting models fit within the memory, compute, and latency budgets of the intended hardware. FLOPs and Ops act as proxies for inference latency and energy consumption \cite{paper21}, while Flash and RAM bounds control the model’s memory footprint.

\subsection{Hardware-Constrained ML Models via Grid Search}


The ML search space includes three tree-based ensemble methods: Light Gradient Boosting Machine (LightGBM), Extreme Gradient Boosting (XGBoost), and Random Forest (RF). These models are chosen for their strong performance on structured traffic features, as reported in prior work \cite{paper28,paper29}, and because, unlike k-NN or SVM whose inference cost scales with the reference set or kernel evaluations, their per-inference cost can be expressed in split comparisons (Ops), which integrates naturally with our hardware-constraint checks.

Hyperparameters are tuned via grid search over the Cartesian product of predefined ranges, using identical train/validation splits and preprocessing across models to ensure a fair and reproducible comparison.

For each configuration explored, hardware measures are analytically estimated from the trained model structure prior to deployment. Ops are derived as an upper-bound from the total number of split comparisons implied by the ensemble’s structure. Flash usage is the estimated memory required to store all decision nodes, and peak RAM usage is computed from the maximum tree depth combined with the memory needed for traversal state and temporary buffers. Each configuration is evaluated for validation accuracy and profiled against these measures; configurations violating any constraint are discarded, and the retained baselines are those achieving the highest validation performance within all budgets.

\subsection{Hardware-Aware Neural Architecture Search (HW-NAS)}

For DL, we use HW-NAS to automatically design compact 1D-CNN architectures optimized for the same hardware budgets. 1D-CNNs are chosen as the backbone since prior work ~\cite{paper24,paper25,paper27} shows their strong performance on network traffic with lower memory and compute needs than RNNs or Transformers. Although based on handcrafted traffic features derived from packet headers and flow metadata, 1D-CNNs effectively capture local correlations that reveal attack patterns while remaining efficient under tight device constraints.



The HW-NAS search space is defined as block-wise 1D-CNN architectures, where each block can vary in number of filters, kernel size, stride, padding type, and dropout rate, and may include optional max or average pooling. Each block also contains batch normalization and activation layers, and the network is formed by stacking these blocks sequentially.

The search process, implemented via an evolutionary algorithm~\cite{paper22,paper24,paper25}, runs for a predefined number of generations, with a fixed number of children generated per generation. Starting from an initial population of random architectures, new candidates are created by mutating the best-performing valid parent from the previous generation. Mutations may add, remove, or modify a CNN block. Each candidate undergoes hardware profiling before training; architectures that exceed any constraint are discarded immediately, while valid ones are retained until the target number of children for that generation is reached. The best-performing valid child in each generation becomes the parent for the next. At the end of the process, the model with the highest validation accuracy across all generations is selected as the final architecture, enabling efficient on-device inference.

\section{Experimental Setup}
\vspace{-0.2cm}
\subsection{Dataset}
This study employs the Edge-IIoTset dataset~\cite{paper2}, a recent benchmark tailored for intrusion detection in IoT and industrial IoT environments. It captures labeled network flows representing both benign activity and 14 distinct attack types, such as DDoS over TCP, SQL injection, man-in-the-middle, and backdoor access. These events span multiple protocol layers, including network, transport, and application.

The dataset is available in two curated formats: a DL subset with 2,218,386 samples and an ML subset with 157,800 samples. Each entry is represented by 63 numerical features extracted from flow-level and protocol-level headers.


We consider three classification schemes, each reflecting a progressively finer level of semantic granularity. The binary setting separates benign traffic from all forms of attack, using a general attack label. The 6-class configuration maps the 14 attack types into five broader categories: denial of service, injection, malware, reconnaissance, and man-in-the-middle. The most detailed setup, the 15-class configuration, preserves all individual attack types as originally labeled. This fine-grained mode serves as the principal classification target throughout our evaluation.


\vspace{-0.05cm}

\subsection{Data Preprocessing}
\vspace{-0.04cm}

Real-time inference on low-power edge devices requires selecting features that are both informative and efficient to extract. We retain 47 features from the original set of 63, excluding those that are unsuitable for real-time extraction, such as payload fields, string-based application-layer attributes, and high-cardinality identifiers. These are removed due to their incompatibility with streaming environments and limited relevance for lightweight deployment.

The retained features are all readily derivable from flow and protocol headers without deep packet inspection. They include indicators such as \textit{tcp.flags}, \textit{udp.port}, \textit{icmp.seq\_le}, and basic flow counters, covering key aspects of the network, transport, and flow layers.

We first remove duplicate entries and rows with missing values. All features are normalized to the [0, 1] range using min–max scaling for numerical stability. The same preprocessing is used consistently across the three classification modes and both ML and DL pipelines, with one-hot encoding of class labels applied during training in the DL case. The complete list of selected features is available in our project repository\footnote{\url{https://github.com/AliiDiabb/Efficient-AI-based-Intrusion-Detection-System-for-IOT-Networks}}.

\vspace{-0.1cm}

\subsection{Target Gateway Platform}

Our deployment target is the Raspberry Pi3 Model~B+, a compact, Linux-based single-board computer intended for use as a gateway in a local network. In this role, it may process local traffic and support additional services such as logging, monitoring, or lightweight control functions.

Table~\ref{tab:pi-specs} lists the key hardware specifications, covering compute capability, memory, storage, connectivity, and power requirements. These features make the Pi suitable for continuous operation in network edge environments where resources must be shared among multiple tasks.

\begin{table}[h]
\centering
\caption{Key specs of the Raspberry~Pi~3 Model~B+.}
\vspace{-0.1cm}
\label{tab:pi-specs}
\begin{tabular}{|c|c|}
\hline
\textbf{Component} & \textbf{Specification} \\\hline
CPU & Quad-core 1.4\,GHz Cortex-A53 \\\hline
RAM & 1\,GB LPDDR2 \\\hline
Storage & microSD (OS and data) \\\hline
Connectivity & Gigabit Ethernet, 802.11ac Wi-Fi, Bluetooth 4.2 \\\hline
OS & Raspbian Lite (Linux) \\\hline
Power & 5\,V, up to 2.5\,A \\ \hline
\end{tabular}
\vspace{-0.3cm}
\end{table}

\subsection{Model Search and Implementation}
Experiments use preprocessed tabular features from the 15-class classification scenario, the primary task of the dataset, and are conducted on a workstation powered by an Nvidia RTX~3070~Ti GPU.

Given the modest resources of the target device (Raspberry~Pi~3~B+)  and the need for stable, low-latency operation alongside other gateway tasks, we set conservative deployment budgets of $\leq 300$~KB flash, $\leq 50$~KB RAM, and $\leq 1.5\times 10^{6}$ operations (Ops for ML, FLOPs for DL). These limits apply uniformly to both ML and DL models and are intentionally well below both the device’s nominal capacity and the resource footprints in related work on tabular traffic data, with the latter’s hardware measures either computed from architectural details or taken directly from reported metrics. This ensures our approach remains robust under varying network loads, avoids resource contention, and leaves ample headroom for other on-device processes.

Table~\ref{tab:search-spaces} summarizes the hyperparameter ranges explored for both the ML grid search and the HW-NAS process in DL. The upper block lists the unified meta-parameters used across the three ensemble classifiers (RF, LightGBM, XGBoost), where each column corresponds to a tunable parameter (e.g., number of trees, maximum depth, feature subsampling ratio) and the second row specifies the discrete values tested; here, \texttt{min\_child\_size} is a generic label mapped to \texttt{min\_samples\_leaf} in RF, \texttt{min\_child\_samples} in LightGBM, and \texttt{min\_child\_weight} in XGBoost (interpreted as the minimum sum of instance weights). The \texttt{num\_leaves} parameter applies only to LightGBM and controls the maximum number of leaves per tree.
The lower block shows the HW-NAS search space for 1D CNN architectures, with each column giving one architectural attribute (e.g., filter count, kernel size, stride, dropout rate, pooling configuration, padding type) and the row beneath indicating the allowed range or options. This layout enables a quick comparison of the tuning scope across tree-based ML models and DL.

\begin{table*}[htb]
\centering
\caption{Search spaces for the ML grid search and HW-NAS for DL}
\vspace{-0.1cm}
\label{tab:search-spaces}
\begin{tabular}{|l|c|c|c|c|c|c|}
\hline
\multicolumn{7}{|c|}{\textbf{Tree-based ML models (unified meta-parameters)}} \\
\hline
\textbf{Param} & $n_{\text{trees}}$ & $\text{max\_depth}$ & $\text{min\_child\_size}$ & $\text{colsample}$ & $\text{subsample}$ & $\text{num\_leaves}$ \\
\hline
\textbf{Range} & $\{2,10,20,50,100,150\}$ & $\{5,10,15,20,25,30\}$ & $\{5,10,20,30,40\}$ & $\{0.6,0.8,1.0\}$ & $\{0.7,0.8,1.0\}$ & $\{8,16,32,64\}$ \\
\hline
\multicolumn{7}{|c|}{\textbf{Deep Learning (HW-NAS, 1D CNN)}} \\
\hline
\textbf{Param} & Filters & Kernel size & Stride & Dropout rate & Pooling & Padding \\
\hline
\textbf{Range} & 16--256 & 2--10 & 1--10 & 0.1--0.5 & max/avg, 2--3 & same/valid \\
\hline
\end{tabular}
\vspace{-0.4cm}
\end{table*}

For ML, the grid search exhaustively evaluates all parameter combinations in the defined space, retaining those that meet the hardware budgets.  
For CNN-1D, HW-NAS runs for 100~generations, producing 15~valid children per generation after hardware check. 
A holdout validation set comprising 20\% of the training data guides architecture selection. Training is performed for up to 100~epochs using the Adam optimizer (initial learning rate~0.008, batch size~2048) and categorical cross-entropy loss, with adaptive learning rate decay when validation loss stagnates, and early stopping to prevent overfitting. 

Models achieving the highest validation accuracy while satisfying all constraints are retrained and assessed across the dataset’s experiments using stratified 5-fold cross-validation. The supporting code is available in our public repository.\textsuperscript{1}

\vspace{-0.05cm}

\section{Results and Discussion}
\label{Results}

\subsection{Multiclass Classification (15-Class Scenario)}
\label{sec:results-multiclass}

We assess model performance on the primary multiclass task, involving 15 attack types, which serves as the target scenario for our grid search and HW-NAS optimization processes. Results are summarized separately for tree-based ML models and DL-based approaches.

Table~\ref{tab:ml_comparison} reports the performance of tree-based ML models on the 15-class task using feature-based inputs. The columns show the average accuracy (Acc.) and F1-score (F1) in \%, flash and RAM usage in KB, and the estimated number of operations (Ops) per inference in thousands. Results for our trained models, shown in bold, are obtained using 5-fold cross-validation with standard deviation; hardware metrics for~\cite{paper2} are not reported due to missing architectural details.

LightGBM achieves the highest accuracy (95.25\%) and F1-score (94.74\%) with the smallest flash (74.93\,KB) and RAM (1.13\,KB) footprint among the tree-based models. XGBoost delivers comparable accuracy (95.11\%) but incurs a flash size of 266.59\,KB and 4.27\,K operations per inference. Random Forest (RF) shows slightly lower accuracy (94.12\%) and F1 (93.21\%), with notably higher RAM usage (4.61\,KB) due to deeper tree traversal. All proposed models significantly outperform the baselines from~\cite{paper2}, which exhibit accuracies between 67.11\% and 80.83\%; this shows our models are effective even in low-compute or resource-constrained tasks.

\begin{table}[h]
\centering
\caption{Comparison with tree-based ML models}
\vspace{-0.2cm}
\label{tab:ml_comparison}
\small
\resizebox{\columnwidth}{!}{%
\scalebox{0.95}{
\begin{tabular}{|>{\centering\arraybackslash}c|c|c|c|c|c|c|}
\hline
\makecell{\textbf{Methods}}  & \makecell{\textbf{Acc.} \\ \textbf{(\%)}} & \makecell{\textbf{F1} \\ \textbf{(\%)}} & \makecell{\textbf{Flash} \\ \textbf{(KB)}} & \makecell{\textbf{RAM} \\ \textbf{(KB)}} & \makecell{\textbf{Ops} \\ \textbf{(K)}} \\
\hline
\textbf{LightGBM} & 
\makecell{\textbf{95.25} \\ \textbf{$\pm$0.13}} &
\makecell{\textbf{94.74} \\ \textbf{$\pm$0.00}} &
\textbf{74.93} & \textbf{1.13} & \textbf{1.20} \\
\textbf{XGBoost} & 
\makecell{\textbf{95.11} \\ \textbf{$\pm$0.12}} &
\makecell{\textbf{ 94.42} \\ \textbf{$\pm$0.00}} &
\textbf{266.59} & \textbf{0.51} & \textbf{4.27} \\
\textbf{RF} & 
\makecell{\textbf{94.12} \\ \textbf{$\pm$0.14}} &
\makecell{\textbf{93.21} \\ \textbf{$\pm$0.00}} &
\textbf{211.22} & \textbf{4.61} & \textbf{3.38} \\
\hline
\makecell{DT~\cite{paper2} \\ RF~\cite{paper2} \\ SVM~\cite{paper2} \\ KNN~\cite{paper2}} &
\makecell{67.11 \\ 80.83 \\ 77.61 \\ 79.18} &
\textemdash & \textemdash & \textemdash & \textemdash \\
\hline
\end{tabular}%
}
\vspace{-0.5cm}
}
\end{table}

Table~\ref{tab:comparison} presents a comparative evaluation between our proposed model (Prop.) and recent DL-based methods. The columns report accuracy (Acc.) and F1-score (F1) (in \%), as well as flash size, peak RAM usage (both in KB), and computational complexity in terms of FLOPs (thousands). All models operate on feature-based inputs. Our results, shown in bold, are obtained using 5-fold cross-validation, with standard deviation shown below each score.

The proposed model achieves 96.73\% accuracy and 97.24\% F1-score with minimal variation, indicating strong performance. While some methods report higher accuracy, such as 99.96\% in~\cite{hassini2024end}, they require significantly more memory and computation, with flash sizes exceeding 1\,MB and FLOPs above 5000\,K. In contrast, our model requires only 190\,KB of flash, 6.89\,KB of RAM, and 838.89\,K FLOPs.

The method proposed in \cite{de2022hybrid} is lightweight in computation with only 52.03\,K FLOPs but suffers from a low F1-score of 74.62\%, suggesting weak class-wise generalization. The applied method in \cite{paper2} reports 94.67\% accuracy, and the proposed method in \cite{aslam2024binary} reports 84.00\% accuracy with 83.00\% F1; both are lower than ours and do not provide architectural details. In contrast, the presented method in \cite{selem2025deep} achieves 98.20\% accuracy but requires over 12\,MB of flash and 8839.18\,K FLOPs, limiting its practicality. Overall, the model proposed in this work offers a favorable balance between performance and efficiency, making it suitable for constrained environments where both accuracy and resource usage matter.

\begin{table}[h]
\centering
\caption{Comparison with DL state-of-the-art methods}
\vspace{-0.2cm}
\label{tab:comparison}
\small
\resizebox{\columnwidth}{!}{%
\scalebox{0.95}{
\begin{tabular}{|>{\centering\arraybackslash}c|c|c|c|c|c|c|}
\hline
\makecell{\textbf{Methods}} & \makecell{\textbf{Acc.} \\ \textbf{(\%)}} & \makecell{\textbf{F1} \\ \textbf{(\%)}} & \makecell{\textbf{Flash} \\ \textbf{(KB)}} & \makecell{\textbf{RAM} \\ \textbf{(KB)}} & \makecell{\textbf{FLOPs} \\ \textbf{(K)}} \\
\hline
\textbf{Prop.} &
\makecell{\textbf{96.73} \\ \textbf{$\pm$0.08}} &
\makecell{\textbf{97.24} \\ \textbf{$\pm$0.07}} &
\textbf{190.34} & \textbf{6.89} & \textbf{838.89} \\
\hline
\cite{de2022hybrid} & 97.14 & 74.62 & 491.52 & 3.60 & 52.03 \\ \hline
\cite{hassini2024end} & 99.96 & 99.00 & 1310.72 & 32.24 & 5651.93 \\ \hline
\cite{ullah2024attention} & 99.95 & 99.60 & 1105.92 & 13.84 & 2971.13 \\ \hline
\cite{selem2025deep}  & 98.20 & -- & 12245.44 & 43.36 & 8839.18 \\ \hline
\cite{paper2} & 94.67 & -- & -- & -- & -- \\ \hline
\cite{aslam2024binary} & 84.00 & 83.00 & -- & -- & -- \\ \hline

\end{tabular}%
}
}
\vspace{-0.3cm}
\end{table}

\subsection{Extended Evaluation Scenarios}

Table~\ref{tab:ml_combined} presents the performance of tree-based ML models on two additional scenarios: a 6-class attack classification task and a 2-class (benign vs malicious) binary task. Each block reports the average accuracy (Acc.), F1-score (F1), flash and RAM usage (in KB), and number of operations (Ops) per inference in thousands. Results for our models are shown in bold and are obtained using 5-fold cross-validation; other works are shown for comparison, though hardware metrics are not reported due to lack of implementation details.


All models maintain satisfactory performance across both tasks, with accuracy above 94\% in the 6-class case and over 99.9\% in the binary setting. LightGBM offers the best trade-off, achieving 99.96\% F1 in the binary task with just 5.02\,KB of flash and 0.08\,K ops. XGBoost remains efficient, while RF incurs slightly higher RAM and operation counts; unlike CNNs with fixed inference cost, ML resource usage scales with the number of output classes. In contrast, baseline models from~\cite{paper2,hamza2023malware} show lower accuracy (77.90–94.10\%), and omit hardware metrics, limiting their deployability comparison.

\label{sec:results-other}
\begin{table}[t]
\centering
\caption{ML models on 6- and 2-class tasks.}
\vspace{-0.2cm}
\label{tab:ml_combined}
\small
\scalebox{0.95}{
\resizebox{\columnwidth}{!}{%
\begin{tabular}{|c|c|c|c|c|c|}
\hline
\textbf{Methods} & \makecell{\textbf{Acc.} \\ \textbf{(\%)}} & \makecell{\textbf{F1} \\ \textbf{(\%)}} & \makecell{\textbf{Flash} \\ \textbf{(KB)}} & \makecell{\textbf{RAM} \\ \textbf{(KB)}} & \makecell{\textbf{Ops} \\ \textbf{(K)}} \\
\hline
\multicolumn{6}{|c|}{\textbf{6-Class Classification}} \\
\hline
\textbf{ LightGBM} & \makecell{\textbf{95.72} \\ \textbf{$\pm$0.12}} & \makecell{\textbf{96.04} \\ \textbf{$\pm$0.00}} & \textbf{30.02} & \textbf{1.13} & \textbf{0.48} \\
\textbf{XGBoost}  & \makecell{\textbf{95.48} \\ \textbf{$\pm$0.13}} & \makecell{\textbf{95.94} \\ \textbf{$\pm$0.00}} & \textbf{110.35} & \textbf{0.51} & \textbf{1.77} \\

\textbf{RF}   & \makecell{\textbf{94.53} \\ \textbf{$\pm$0.15}} & \makecell{\textbf{94.98} \\ \textbf{$\pm$0.02}} & \textbf{189.96} & \textbf{4.61} & \textbf{3.03} \\
\hline

\makecell{DT~\cite{paper2} \\ RF~\cite{paper2} \\ SVM~\cite{paper2} \\ KNN~\cite{paper2}} & 
\makecell{77.90 \\ 82.90 \\ 85.62 \\ 83.39} & 
\textemdash & \textemdash & \textemdash & \textemdash \\

\hline
\makecell{KNN~\cite{hamza2023malware} \\ DT~\cite{hamza2023malware} \\ LR~\cite{hamza2023malware} \\ SVM~\cite{hamza2023malware} \\ RF~\cite{hamza2023malware}} & 
\makecell{90.30 \\ 92.50 \\ 81.50 \\ 84.30 \\ 94.10} & 
\makecell{90.30 \\ 97.50 \\ 81.50 \\ 84.30 \\ 94.00} & 
\textemdash & \textemdash & \textemdash \\
\hline
\multicolumn{6}{|c|}{\textbf{2-Class (Binary) Classification}} \\
\hline
\textbf{LightGBM} & \makecell{\textbf{99.93} \\ \textbf{$\pm$0.00}} & \makecell{\textbf{99.96} \\ \textbf{$\pm$0.00}} & \textbf{5.02} & \textbf{1.13} & \textbf{0.08} \\

\textbf{XGBoost}  & \makecell{\textbf{99.91} \\ \textbf{$\pm$0.01}} & \makecell{\textbf{99.95} \\ \textbf{$\pm$0.01}} & \textbf{18.22} & \textbf{0.51} & \textbf{0.29} \\

\textbf{RF}   & \makecell{\textbf{99.92} \\ \textbf{$\pm$0.03}} & \makecell{\textbf{99.94} \\ \textbf{$\pm$0.00}} & \textbf{6.25} & \textbf{1.84} & \textbf{0.1} \\
\hline

\makecell{DT~\cite{paper2} \\ RF~\cite{paper2} \\ SVM~\cite{paper2} \\ KNN~\cite{paper2}} & 
\makecell{99.98 \\ 99.99 \\ 99.99 \\ 99.99} & 
\textemdash & \textemdash & \textemdash & \textemdash \\

\hline

\end{tabular}
}
}
\vspace{-0.1cm}

\end{table}

Table~\ref{tab:dl_extended} reports the performance of DL models on the 6-class and 2-class tasks. The columns include accuracy (Acc.), F1-score (F1), flash and RAM usage (in KB), and FLOPs (K); our results are in bold and use 5-fold cross-validation.

The proposed model (Prop.) achieves 96.83\% accuracy on 6-class and 99.74\% on the binary task, with modest memory (190.34\,KB flash, 6.89\,KB RAM) and compute cost (838.89\,K FLOPs). \cite{de2022hybrid} is lighter (52.03\,K FLOPs) but less accurate (97.85\%). \cite{aslam2024binary} performs well on binary but generalizes weakly to 6-class. \cite{paper2} reports high binary accuracy (99.99\%) but lower 6-class score (96.01\%) and omits F1 and hardware metrics, limiting comparison. Overall, our model balances performance and efficiency for practical deployment.


\begin{table}[t]
\centering
\caption{DL models on 6-class and 2-class tasks}
\vspace{-0.09cm}
\label{tab:dl_extended}
\huge
\resizebox{\columnwidth}{!}{%
\scalebox{0.95}{
\begin{tabular}{|c|c|c|c|c|c|c|c|c|}
\hline
\multirow{2}{*}{\textbf{Methods}} & \multicolumn{2}{c|}{\textbf{6-Class}} & \multicolumn{2}{c|}{\textbf{2-Class}} & \multirow{2}{*}{\makecell{\textbf{Flash} \\ \textbf{(KB)}}} & \multirow{2}{*}{\makecell{\textbf{RAM} \\ \textbf{(KB)}}} & \multirow{2}{*}{\makecell{\textbf{FLOPs} \\ \textbf{(K)}}} \\ \cline{2-5}
 & \makecell{\textbf{Acc.} \\ \textbf{(\%)}} & \makecell{\textbf{F1} \\ \textbf{(\%)}} & \makecell{\textbf{Acc.} \\ \textbf{(\%)}} & \makecell{\textbf{F1} \\ \textbf{(\%)}} & & & \\ 
\hline
\textbf{Prop.} &
\makecell{\textbf{96.83} \\ \textbf{$\pm$0.06}} &
\makecell{\textbf{96.88} \\ \textbf{$\pm$0.05}} &
\makecell{\textbf{99.74} \\ \textbf{$\pm$0.02}} &
\makecell{\textbf{99.74} \\ \textbf{$\pm$0.02}} &
\textbf{190.34} & \textbf{6.89} & \textbf{838.89} \\

\hline
\cite{paper2} & 96.01 & -- & 99.99 & -- & -- & -- & -- \\
\hline

\cite{de2022hybrid} & -- & -- & 97.85 & 97.81 & 491.52 & 3.60 & 52.03\\
\hline
\cite{aslam2024binary} & 84.00 & 83.00 & 100.00 & 100.00 & -- & -- & -- \\
\hline

\end{tabular}%
}}
\vspace{-0.4cm}
\end{table}

\vspace{-0.2cm}

\section{Edge Deployment and Evaluation}
\label{Deployment}
\subsection{Deployment Setup}

We evaluate the on-device deployment performance of the trained classifiers on a Raspberry~Pi~3~B+, configured as a local edge gateway for IoT/IIoT traffic classification. In this setup, the gateway captures network traffic in real time, performs packet-level sniffing, and aggregates packets into unidirectional flows between two endpoints based on five identifiers: source and destination IP addresses, source and destination ports, and transport-layer protocol. From each flow, the 47 structured features used in this study are extracted directly on the device; inference is then performed locally without quantization or compression. The CNN operates in float32 precision, while the tree-based ML models run using their default inference formats.

Power and latency measurements were collected using a USB power meter placed between the Raspberry~Pi and its power supply. The device records current draw at high resolution during each prediction; we compute the instantaneous power as \(P = I \times V\), and derive the energy per inference as the product of average power and measured latency.

\subsection{On-Device Inference Efficiency}

All reported values correspond to the additional consumption during inference, measured above the idle baseline of 500\,mA (2.5\,W). Feature extraction, executed on-device immediately before each prediction, averaged 0.93\,ms across all models and is excluded from the reported latency below.

Table~\ref{tab:deployment_results} reports the average inference latency, current draw, power consumption, and energy per prediction for each model on the Raspberry~Pi~3~B+; values are averaged over repeated runs and expressed respectively in milliseconds (ms), milliamperes (mA), milliwatts (mW), and millijoules (mJ).

\begin{table}[h!]
\centering
\caption{On-device performance of classifiers.}
\vspace{-0.1cm}
\label{tab:deployment_results}
\scriptsize
\resizebox{\columnwidth}{!}{%
\begin{tabular}{|l|c|c|c|c|}
\hline
\makecell{\textbf{Model}} & 
\makecell{\textbf{Latency} \\ \textbf{(ms)}} & 
\makecell{\textbf{Current} \\ \textbf{(mA)}} & 
\makecell{\textbf{Power} \\ \textbf{(mW)}} & 
\makecell{\textbf{Energy} \\ \textbf{(mJ)}} \\ \hline

Random Forest & 70   & 65 & 325 & 22.75 \\\hline
LightGBM      & 27   & 50 & 250 & 6.75  \\\hline
XGBoost       & 22   & 50 & 250 & 5.50  \\\hline
CNN           & 300  & 55 & 275 & 82.50 \\\hline
\end{tabular}
}
\vspace{-0.1cm}
\end{table}

LightGBM and XGBoost yield the lowest energy footprints, under 7\,mJ per prediction, while maintaining sub-30\,ms latency; this makes them suitable for deployments with tight energy and response-time constraints. The CNN model, while more accurate, incurs over 82\,mJ per prediction due to its longer runtime and floating-point operations, which makes it better suited for scenarios where higher accuracy is desired and latency can be moderately tolerated. Random Forest exhibits slightly higher energy than the boosting models despite a lower operation count, likely due to less efficient memory access patterns during tree traversal.

\section{Conclusion}
\label{Conclusion}
\vspace{-0.1cm}

This work shows that accurate intrusion detection can be achieved without compromising hardware feasibility; it aligns model design with platform constraints while exposing key accuracy–efficiency trade-offs. Tree-based ensembles, particularly LightGBM, offer strong performance with minimal resource usage, making them ideal for deployment on constrained edge devices. HW-NAS optimized 1D-CNNs further improve accuracy with moderate cost and remain practical for edge inference without overloading the system. Both models minimize overhead and suit a wide range of embedded or gateway-class devices, as they occupy only a small fraction of memory and compute resources. Empirical results confirm that the selected models meet deployment constraints across tasks, while outperforming recent deep baselines in efficiency and remaining competitive in accuracy. Future directions include compressing trained models (e.g., quantization, pruning), incorporating explainability techniques, and evaluating robustness under adversarial or noisy traffic.



\end{document}